%% file: extrapolate_main.tex
\documentclass[sigconf,natbib=true,anonymous=False]{acmart}
\usepackage{graphicx}
\usepackage{multirow}
\usepackage{bm}
\usepackage{mwe}
\usepackage{balance}
\usepackage{enumitem}
\usepackage{caption}
\usepackage{amsthm}
\usepackage{amsmath}
\usepackage{booktabs}
\usepackage{subfig}
\usepackage{siunitx}
\usepackage{footmisc}
\usepackage{footnote}
\usepackage[flushleft]{threeparttable}
\usepackage{etoolbox}
\usepackage{array}

\appto\TPTnoteSettings{\footnotesize}

\usepackage{tablefootnote}
\makesavenoteenv{table}
\makesavenoteenv{tabular}

\newcolumntype{C}[1]{>{\centering\arraybackslash}p{#1}}
\newcolumntype{R}[1]{>{\raggedleft\arraybackslash}p{#1}}
\newcolumntype{L}[1]{>{\raggedright\arraybackslash}p{#1}}

\newcommand\mc[1]{\multicolumn{1}{c}{#1}}

\newcommand\blfootnote[1]{%
\begingroup 
\renewcommand\thefootnote{}\footnote{#1}%
\addtocounter{footnote}{-1}%
\endgroup 
}

\makeatletter
\newcommand{\ssymbol}[1]{^{\@fnsymbol{#1}}}
\makeatother

\makeatletter
\def\hlinew#1{%
  \noalign{\ifnum0=`}\fi\hrule \@height #1 \futurelet
   \reserved@a\@xhline}
\makeatother

\AtBeginDocument{%
  \providecommand\BibTeX{{%
    \normalfont B\kern-0.5em{\scshape i\kern-0.25em b}\kern-0.8em\TeX}}}

\setcopyright{acmcopyright}
\copyrightyear{2022}
\acmYear{2022}
\acmDOI{}
\acmISBN{}

\acmConference[CIKM '22]{Proceedings of the 31st ACM International Conference on Information and Knowledge Management}{October 17--22, 2022}{Atlanta, Georgia, USA.}

\begin{document}

\title{Evaluating Interpolation and Extrapolation Performance of Neural Retrieval Models}

%
\author{Jingtao Zhan$^{1}$, Xiaohui Xie$^{1}$, Jiaxin Mao$^{2}$, Yiqun Liu$^{1\star}$, Jiafeng Guo$^{3}$, Min Zhang$^{1}$, Shaoping Ma$^{1}$}
\affiliation{%
  \institution{${1}$ Department of Computer Science and Technology, Beijing National Research Center for Information Science and Technology, Tsinghua University, Beijing 100084, China
  }
  \country{}
  }
  
\affiliation{%
  \institution{${2}$ Beijing Key Laboratory of Big Data Management and Analysis Methods, Gaoling School of Artificial Intelligence, \\ Renmin University of China, Beijing 100872, China
}
  \country{}
  }
  
  \affiliation{%
  \institution{${3}$ CAS Key Lab of Network Data Science and Technology, Institute of Computing Technology, \\
  		Chinese Academy of Sciences, Beijing, China
  }
  \country{}
  }

\email{jingtaozhan@gmail.com, yiqunliu@tsinghua.edu.cn}

\renewcommand{\shortauthors}{Zhan, et al.}

\begin{abstract}
A retrieval model should not only \emph{interpolate} the training data but also \emph{extrapolate} well to the queries that are different from the training data. While neural retrieval models have demonstrated impressive performance on ad-hoc search benchmarks, we still know little about how they perform in terms of interpolation and extrapolation. In this paper, we demonstrate the importance of separately evaluating the two capabilities of neural retrieval models. 
Firstly, we examine existing ad-hoc search benchmarks from the two perspectives. We investigate the distribution of training and test data and find a considerable overlap in query entities, query intent, and relevance labels. This finding implies that the evaluation on these test sets is biased toward interpolation and cannot accurately reflect the extrapolation capacity. 
Secondly, we propose a novel evaluation protocol to separately evaluate the interpolation and extrapolation performance on existing benchmark datasets. It resamples the training and test data based on query similarity and utilizes the resampled dataset for training and evaluation. 
Finally, we leverage the proposed evaluation protocol to comprehensively revisit a number of widely-adopted neural retrieval models. Results show models perform differently when moving from interpolation to extrapolation. For example, representation-based retrieval models perform almost as well as interaction-based retrieval models in terms of interpolation but not extrapolation. Therefore, it is necessary to separately evaluate both interpolation and extrapolation performance and the proposed resampling method serves as a simple yet effective evaluation tool for future IR studies.
\end{abstract}

\ccsdesc[500]{Information systems~Retrieval effectiveness}
\ccsdesc[500]{Information systems~Retrieval models and ranking}

\keywords{evaluation, extrapolation, interpolation, neural ranking}

\maketitle

\blfootnote{$^\star$Corresponding author}

\input{introduction.tex}

\input{Definition}

\input{datasets}
\input{test_train_overlap}

\input{resample_method}


\input{cmp_architecture}
\input{cmp_dr_train}

\input{related_work}

\input{conclusions}

\begin{acks}
This work is supported by the Natural Science Foundation of China (Grant No. 61732008) and Tsinghua University Guoqiang Research Institute.
\end{acks}

\bibliographystyle{ACM-Reference-Format}
\newpage
\balance
\bibliography{references}

\end{document}

%% file: introduction.tex
\section{Introduction}

Ranking is essential for many IR-related tasks, such as ad-hoc search~\cite{zhan2021optimizing, xiong2021approximate}, Web search~\cite{liu2021pre}, and open domain question answering~\cite{guu2020realm, karpukhin2020dense}. 
Recently, there has been a surge of research interest in applying neural models for first-stage retrieval to improve ranking performance considering their strong capacity for semantic matching. 
Various model architectures~\cite{Khattab2020ColBERTEA, karpukhin2020dense, lin2021few, formal2021splade, deepimpact} and training techniques~\cite{xiong2021approximate, zhan2021optimizing, qu2021rocketqa, lin2020distilling, hofstatter2021efficiently, gao2021unsupervised, lu2021less} are proposed. 
They are typically evaluated and compared on ad-hoc retrieval benchmarks such as MS MARCO~\cite{bajaj2016ms} and TREC Deep Learning Tracks~\cite{craswell2019overview, Craswell2020Overview}.
The evaluation protocol is to train the models with massive annotation labels and then evaluate them on a held-out test set. 
Results suggest that neural retrieval models strongly outperform traditional term-based retrieval methods like BM25~\cite{bm25} and drive test-set performance to new heights~\cite{qu2021rocketqa, karpukhin2020dense, gao2021unsupervised, zhan2022learning}.
 
However, an overall evaluation metric score on a dataset with unclear training-test distribution does not show the full picture of model performance. In this paper, we break the model capacity into interpolation and extrapolation regimes, which are two fundamental concepts in machine learning researches~\cite{barnard1992extrapolation, balestriero2021learning, haley1992extrapolation, xu2020neural, webb2020learning}. Because queries play a central role in ad-hoc search, we \emph{define the interpolation performance for ranking as how ranking models handle test queries that are similar to training queries}. Similarly, \emph{the extrapolation performance is defined as how ranking models perform on test queries that are distinct from training queries}. It is commonly believed that extrapolation is a more intricate task than interpolation~\cite{balestriero2021learning, rosenfeld2022online}. It is also a more important task for neural retrieval models since retrieval systems should be able to handle arbitrary search requests.
Nevertheless, despite the impressive benchmark performance scores of neural retrieval models, we know relatively little about how they perform in interpolation and extrapolation scenarios, respectively. 

We first try to analyze and interpret test-set scores on the benchmarks from the interpolation and extrapolation perspectives in Section~\ref{sec:investigate_benchmark}.
We investigate the distribution of training and test data on TREC Deep Learning Tracks~\cite{craswell2019overview, Craswell2020Overview} and MS MARCO dataset~\cite{bajaj2016ms}, both of which are popular ranking benchmarks~\cite{Craswell2021MSMB}. 
The investigation leads to an alarming finding that a substantial training-test overlap exists in terms of query entities, query intents, and relevance labels.
Therefore, the existing benchmarks largely evaluate the interpolation ability but fail to accurately reflect the extrapolation ability.

Second, we propose a new evaluation protocol to evaluate the interpolation and extrapolation performance based on an existing Cranfield-like dataset in Section~\ref{sec:resampling_evaluation}. We encode the queries in an embedding space and use embedding distance as query similarity. Then we resample similar or dissimilar training-test data to construct a new dataset for the evaluation of the interpolation or extrapolation performance, respectively. Models are trained and evaluated on the new dataset. 
Note that the proposed evaluation method does not require any new test datasets or relevance annotation. To investigate whether it is limited to the distribution that the original dataset follows, we investigate its alignment with generalization performance evaluated by a new out-of-distribution test set. Results suggest that extrapolation performance strongly correlates to out-of-distribution performance while interpolation correlates poorly. Therefore, the proposed evaluation protocol also indicates generalization performance beyond the dataset used for resampling. 

Finally, we utilize the proposed evaluation protocol to comprehensively revisit various neural retrieval models in Section~\ref{sec:evaluate_extrapolation}. We study whether the evaluation results in terms of interpolation and extrapolation lead to different outcomes in model comparisons. We evaluate different model architectures including interaction-based models~(ColBERT~\cite{Khattab2020ColBERTEA}) and representation-based models~(dense retrieval~\cite{karpukhin2020dense} and neural sparse retrieval~\cite{formal2021splade}). We also evaluate some popular training techniques including hard negative mining~\cite{xiong2021approximate, zhan2021optimizing}, distillation~\cite{hofstatter2021efficiently, lin2020distilling}, and pretraining~\cite{gao2021condenser, gao2021unsupervised}. 
Extensive experimental comparisons are performed. Results demonstrate a substantial disparity in model effectiveness evaluated in the two regimes.
For example, interaction-based models extrapolate well, whereas dense retrieval and neural sparse retrieval suffer from severe effectiveness degeneration when moving from interpolation to extrapolation. Besides, hard negative mining and distillation hardly mitigate the performance gap between the two regimes, while the gap is tightly related to the pretraining technique. Therefore, it is necessary to evaluate the interpolation and extrapolation capacities separately to gain a deep understanding of model performance\footnote{
Code is available at \url{https://github.com/jingtaozhan/extrapolate-eval}.
}.

%% file: Definition.tex
\section{Interpolation And Extrapolation}
\label{sec:define_interpolation_extrapolation}

Interpolation and extrapolation are fundamental concepts in characterizing the behavior of machine learning algorithms on different test cases. 
While there is no widely-accepted precise formulation, an intuitive and informal definition is commonly agreed~\cite{barnard1992extrapolation, barbiero2020modeling, webb2020learning}: interpolation occurs when the training set is fully representative of the test data, while extrapolation occurs when the training data differs distinctly from the test data. For example, if a machine-learning algorithm is required to apply arithmetics to much larger numbers than those seen during training, it has to extrapolate. Humans extrapolate well in many scenarios because the learning process is beyond simple pattern recognition and that humans are able to generalize knowledge to new situations~\cite{lake2017building}. On the contrary, machine learning algorithms excel at interpolation but may be vulnerable to extrapolation~\cite{haley1992extrapolation, barnard1992extrapolation}. To achieve high extrapolation accuracy, the algorithms have to learn faithful knowledge that generally holds for all possible inputs instead of spurious correlations that only work for most training data. Evaluating the extrapolation performance is of utmost importance to building trustworthy and robust systems that can be applied in the real world. It has fostered a bunch of machine-learning research in exploring generalizable neural networks~\cite{xu2020neural, haley1992extrapolation, barnard1992extrapolation} and designing robust optimization algorithms~\cite{krueger2021out, sagawa2019distributionally, webb2020learning}. 

Although ranking is an application area of machine learning, few studies consider interpolation and extrapolation when evaluating ranking models. Since the Cranfield experiments~\cite{cleverdon1966aslib}, evaluation studies in IR focus on measuring the effectiveness of search systems on a test collection (i.e. a test set for ranking models) but largely overlook the training process of ranking models and the relationship between training and test data. A typical IR test collection consists of three major components: a corpus of texts to be retrieved, a set of test queries, and a set of relevance labels for some query-text pairs. To evaluate the effectiveness of a ranking model, we can compute some evaluation metrics for each test query with the relevance labels. Based on different assumptions about user behavior, plenty of effort has been put to propose evaluation metrics that reflect user satisfaction, such as NDCG~\cite{jarvelin2002cumulated}, RBP~\cite{moffat2008rank}, INST~\cite{bailey2015user}, BPM~\cite{zhang2017evaluating}, IFT~\cite{azzopardi2018measuring}, etc. 
While such test-collection-based evaluations have been widely adopted in evaluating ranking models, most existing test collections do not contain any training data. With the emergence of neural IR models, some recent test collections, such as MS MARCO~\cite{bajaj2016ms}, TREC Deep Learning Tracks~\cite{Craswell2020Overview, craswell2019overview}, and NTCIR-WWW2\&3~\cite{sakai2020overview, mao2019overview}, provide large-scale training sets to facilitate the employment of neural ranking models. However, few researchers and practitioners have considered how the training data may impact the performance on test queries, which is especially important in the large data regime where the superior performance of neural ranking models originates from massive training queries. 

In this paper, we argue that interpolation and extrapolation are two important perspectives for a proper evaluation of the ranking models that need to be trained with large-scale training data. A model with strong interpolation performance does not necessarily extrapolate well. If the ranking models learn misleading heuristics about relevance that are only suitable for training data, they may still interpolate well on similar inputs but extrapolate substantially worse. Therefore, high metric scores on interpolation test collections may not indicate high user satisfaction when the models are applied in the real world where extrapolation is the norm.

Now we consider how to define interpolation and extrapolation in IR. Note that machine-learning researchers usually agree on an intuitive and informal definition that interpolation and extrapolation are the cases where test points are similar to or considerably different from the training data, respectively~\cite{barnard1992extrapolation, barbiero2020modeling, webb2020learning}. As for precise definitions, there are ongoing arguments about what definitions better align with the generalization performance\footnote{Some researchers formally define interpolation and extrapolation as whether the test points lie within or outside the convex hull bounded by the extremes of the training data~\cite{rosenfeld2022online, webb2020learning, barbiero2020modeling, krueger2021out}. However, such geometric definition has been challenged by recent studies~\cite{balestriero2021learning, zeni2022exploring} which show that it is uninformative for the case of high-dimensional spaces and poorly aligns with the generalization performance.}. In this paper, we do not address the problem of mathematically formalizing the two concepts. Instead, we explore a practical definition that characterizes the generalization performance of ranking models when deployed in real-world search engines. In the context of ad-hoc search, queries are of utmost importance to both users and ranking systems. Queries are abstractions of the user intent, and the ranking systems rely on queries to effectively retrieve relevant items that satisfy the users. Queries are not static but dynamically changing: query formulations differ among users, user intents vary across domains, and user interest shifts over time. An effective ranking system should be able to deal with any possible query no matter whether it is similar to the training queries or not. Therefore, we propose to define interpolation and extrapolation in ad-hoc search as whether the test query is similar to the training queries. 
 Concretely, our definitions are:
\begin{itemize}
	\item Interpolation occurs for a test query when there is at least one similar or duplicate training query. 
	\item Extrapolation occurs for a test query when there is no similar or duplicate training query.
\end{itemize}
We acknowledge that there can be different implementations of computing query similarity. For example, it can be annotated in terms of intents, entities, etc. It can also be computed automatically based on the numbers of matched terms or embedding distance. 

Our exploration sheds some light on this problem and helps inspire future studies in this direction. 
In light of the important role of queries in ad-hoc search, queries are exclusively considered in our definitions of interpolation and extrapolation. But we also note that documents can also be integrated into the definitions. In modern search engines, not only queries but also documents are dynamically changing. New documents are constantly indexed, and the ranking systems should be capable of generalizing to them which may differ distinctly from the old ones. In our study, we view the corpus as a static set and only consider the dynamics of queries. Further incorporating the dynamics of documents may better characterize the generalization performance when the ranking models are applied in practice. We will explore it in the future.

%% file: datasets.tex
\section{Investigating Existing Benchmarks}
\label{sec:investigate_benchmark}

After defining interpolation and extrapolation in IR, now we utilize them to investigate the ad-hoc search benchmarks that are popular in the large data regime of neural IR. MS MARCO dataset~\cite{bajaj2016ms} and TREC Deep Learning Tracks~(TREC DL)~\cite{craswell2019overview, Craswell2020Overview} are widely-adopted benchmarks for training and evaluating neural ranking models~\cite{Craswell2021MSMB}. They consist of massive training queries that strongly benefit neural ranking models. Because most neural ranking models since the BERT era~\cite{devlin2019bert} are evaluated and compared on the benchmarks, \citet{Lin2021PretrainedTF} even point out that the rapid progress in IR would not have been possible without them. We will introduce the dataset details and then investigate whether the test-set performance reflects interpolation or extrapolation capacity.

\subsection{Benchmark Details}
\label{sec:datasets_marco_trec}

MS MARCO~\cite{bajaj2016ms} and TREC DL~\cite{craswell2019overview, Craswell2020Overview} consist of two tightly-related tasks: passage ranking and document ranking, which share very similar training and test queries. MS MARCO transfers passage ranking labels to the document ranking task. TREC DL shares the same test queries between the two tasks. To save the cost of annotation and computation, this paper only focuses on the passage ranking task. But the findings also generalize to the document ranking task due to their tight relation. In the following, we elaborate on the statistics of the passage ranking task.

MS MARCO and TREC DL share the same corpus and training queries. The corpus has approximately 8.8 million passages extracted from Web pages. The training queries are 0.5 million natural-language queries gathered from the Bing search engine’s log. They are shallowly annotated. On average, only one passage per query is marked relevant.

MS MARCO and TREC DL provide different test queries. 
MS MARCO has a development~(dev) set and an evaluation set. The latter is hidden and is used for the leaderboard. Thus this paper uses the public dev set for investigation.
It has 6,980 natural-language queries sampled from Bing's search log. They are shallowly annotated, and most queries only have one relevant passage. The official evaluation metric is MRR@10. 
As for TREC DL, it provides small test query sets. The query-passage pairs are judged on a four-point scale. A large annotation pool is constructed to identify a sufficiently comprehensive set of relevant passages. We use the query sets in the year 2019 and 2020. They have 43 and 54 test queries, respectively. The official evaluation metric is NDCG@10.

%% file: test_train_overlap.tex
\subsection{Problem of Existing Benchmarks}
\label{sec:test_train_overlap}

To investigate whether MS MARCO and TREC DL evaluate the interpolation or extrapolation performance, we examine query similarity between the training and test sets.
The examination is carried out in three aspects, i.e., whether two queries share similar entities, similar intents, or the same relevant passages.


Similarity in terms of query entities and intents is examined based on manual annotation. 
We first use BM25~\cite{bm25, yang2018anserini} and an embedding model~\cite{zhan2021optimizing} to separately recall ten training queries for each test query. 
Then, three annotators are recruited to label the similarity scores between each test query and the recalled training queries. 
Judgments for entity overlap are on a three-point scale:
\begin{itemize}
	\item \textbf{Full Overlap}: There exists at least one training query that shares exactly the same set of entities with the test query.
	\item \textbf{Partial Overlap}: There exists at least one training query that shares at least one entity with the test query but no training query that fully overlaps with the test query. 
	\item \textbf{Zero Overlap}: All recalled training queries do not share any entity with the test query.
\end{itemize}
Judgments for intent similarity are on a three-point scale:
\begin{itemize}
	\item \textbf{Duplicate}: There exists at least one training query with exactly the same intent as the test query.
	\item \textbf{Similar}: There exists at least one training query with a similar intent as the test query but no duplicate query.
	\item \textbf{Dissimilar}: All recalled training queries are different in intent from the test query.
\end{itemize}
The average Cohen’s Kappa values~\cite{Cohen1960ACO} are 0.59 and 0.54 for entity and intent annotation, respectively. When there is disagreement among the annotators, we use the median score as the final label.  

\input{latex_tables/query_entitiy_overlap.tex}

\input{latex_tables/query_intent_overlap.tex}

Annotation results of entity and intent similarity are shown in Table~\ref{tab:query_entity_overlap} and Table~\ref{tab:query_intent_overlap}, respectively. 
According to Table~\ref{tab:query_entity_overlap}, we can see that more than $90\%$ test query entities are fully or partially overlapped by one training query. Hence the benchmark is capable of evaluating how ranking models interpolate to previously seen query entities but fails to evaluate how ranking models extrapolate to unseen entities.
According to Table~\ref{tab:query_intent_overlap}, we observe a surprisingly high overlap in query intent. More than $50\%$ queries share duplicate or similar intent with training queries. Since we only recall not more than 20 candidates for each test query, the similar training queries for some test queries are missing from the candidates and the true overlap proportions are expected to be higher. Among the three test collections, TREC 19 DL test set is most similar to the training set, where only $21\%$ of test queries are novel. Therefore, all three test collections are biased towards interpolation in terms of intent, and TREC 19 DL is the most biased test set.

\input{latex_tables/relevant_passage_overlap.tex}

After showing the annotation results about query entities and intents, we further investigate the training-test query similarity based on overlap in relevant passages. The benchmark corpus consists of short passages that are on average 56 words long. Considering the limited information a short passage contains, two queries that share the same relevant passage are likely to be similar. Moreover, such overlap in relevant passages also leaves the possibility that ranking models achieve high test-set performance via memorizing training labels. 
Now we examine to what extent the test queries of TREC 19 \& 20 DL test collections~\cite{craswell2019overview, Craswell2020Overview} share the same relevant passages with the training queries. Here we focus on TREC DL test sets instead of MS MARCO dev set because it is shallowly annotated and there is little overlap.
We compute the overlap percentage as follows. We mark a test query similar if its any relevant passage is labeled relevant in the training set. Since TREC 19 \& 20 DL test collections consist of four-point scaled relevance labels, we convert the labels to binary using different thresholding values. 
We show the proportion of overlapped test queries in Table~\ref{tab:relevant_passage_overlap}. It is striking that near $80\%$ queries have at least one somewhat/highly/perfectly relevant passage that is also labeled relevant in the training set. Such a considerable overlap implies strong similarity between training and test queries and further validates that the two TREC DL test collections are strongly biased towards interpolation evaluation.

\input{latex_tables/duplicate_query_example.tex}

Finally, we show several duplicate training-test query pairs.
The test queries are manually selected from the TREC 20 DL test set. A subset of duplicate pairs is shown in Table~\ref{tab:duplicate_query_example}. We can see that the training query is very similar to the test query in terms of intents and entities. Most query formulations are alarmingly similar, which only differ in stopwords or minor changes of the word sequence. If a ranking model performs well on these test queries, it may not extrapolate well when applied in the real world where queries are not necessarily similar to the training set. Therefore, it is important to consider interpolation and extrapolation when evaluating a model on a Cranfield-like dataset, especially nowadays when massive data is used for training neural models.

%% file: latex_tables/query_entitiy_overlap.tex
\begin{table}[t]
    \small
    \caption{
    Distribution of test queries in terms of entity overlap with training queries. The number shows the percentage of test queries that have exactly/partially/zero (the) same entities with training queries. Dev-samples are 100 queries randomly sampled from MS MARCO dev set.}
    \label{tab:query_entity_overlap}
    \begin{tabular}{l|ccc}
    \toprule
    {\textbf{Entity Overlap}} 	& {\textbf{TREC 19 DL}} & {\textbf{TREC 20 DL}} & {\textbf{Dev-samples}}  \\
    \midrule
    {\textbf{Full overlap}} 	& 63\% & 54\% & 45\% \\
    {\textbf{Partial overlap}} 	& 30\% & 39\% & 51\% \\
    {\textbf{Zero overlap}} 		& 7\%  & 7\% & 4\% \\  
    \bottomrule
    \end{tabular}
    \end{table}

%% file: latex_tables/query_intent_overlap.tex
\begin{table}[t]
    \small
    \caption{Distribution of test queries in terms of intent similarity with training queries. The number shows the percentage of test queries that share duplicate/similar/dissimilar intent with training queries. Dev-samples are 100 queries randomly sampled from MS MARCO dev set.}
    \label{tab:query_intent_overlap}
    \begin{tabular}{l|ccc}
    \toprule
    {\textbf{Intent Similarity}} 	& {\textbf{TREC 19 DL}} & {\textbf{TREC 20 DL}} & {\textbf{Dev-samples}}  \\
    \midrule
    {\textbf{Duplicate}} 	& 51\% & 37\% & 34\% \\
    {\textbf{Similar}} 		& 28\% & 22\% & 19\% \\
    {\textbf{Dissimilar}} 	& 21\% & 41\% & 47\% \\    
    \bottomrule
    \end{tabular}
    \end{table}

%% file: latex_tables/relevant_passage_overlap.tex
\begin{table}[t]
    \small
    \caption{Distribution of test queries that share relevant passages with training queries. The numbers show the percentages of test queries that share at least one relevant (somewhat relevant, highly relevant, or perfectly relevant) result with another query in the training set.
    }
    \label{tab:relevant_passage_overlap}
    \begin{tabular}{l|ccc}
    \toprule
    {\textbf{Passage Relevance}} 			& {\textbf{TREC 19 DL}} & {\textbf{TREC 20 DL}}  \\
    \midrule
    {$=3$ \textbf{ Perfectly relevant}} 	& 26\% & 31\%  \\
    {$\geq 2$ \textbf{ Highly relevant}} 	& 60\% & 46\%  \\
    {$\geq 1$ \textbf{ Somewhat relevant}}  & 79\% & 76\%  \\
    \bottomrule
    \end{tabular}
    \end{table}

%% file: latex_tables/duplicate_query_example.tex
\begin{table}[t]
    \small
    \caption{Examples of duplicate test and training queries.}
    \label{tab:duplicate_query_example}
    \begin{tabular}{l|l}
    \toprule 
    \multicolumn{1}{c|}{\textbf{TREC 20 DL Test Query}}	& \multicolumn{1}{c}{\textbf{Duplicate Training Query}} \\
    \midrule 
    \begin{tabular}[l]{p{0.19\textwidth}} average wedding dress alteration cost  \end{tabular} 	
    &  \begin{tabular}[l]{p{0.19\textwidth}} average cost for wedding dress alterations \end{tabular}   \\ \midrule
    \begin{tabular}[l]{p{0.19\textwidth}} what does it mean if your tsh is low  \end{tabular} 	
    & \begin{tabular}[l]{p{0.19\textwidth}} what does it mean if my tsh is low \end{tabular} \\ \midrule
    \begin{tabular}[l]{p{0.19\textwidth}} meaning of shebang  \end{tabular} 	
    & \begin{tabular}[l]{p{0.19\textwidth}} what is a shebang \end{tabular} \\ \midrule
    \begin{tabular}[l]{p{0.19\textwidth}} do google docs auto save  \end{tabular} 	
    &  \begin{tabular}[l]{p{0.19\textwidth}} does google docs automatically save \end{tabular} \\ \midrule
    \begin{tabular}[l]{p{0.19\textwidth}} how long does it take to remove wisdom tooth  \end{tabular} 	
    &  \begin{tabular}[l]{p{0.19\textwidth}} how long does it take to have wisdom teeth removed \end{tabular} \\ \midrule
    \begin{tabular}[l]{p{0.19\textwidth}} how much would it cost to install my own wind turbine  \end{tabular} 	
    &  \begin{tabular}[l]{p{0.19\textwidth}} how much does it cost to install a wind turbine \end{tabular} \\ \midrule
    \begin{tabular}[l]{p{0.19\textwidth}} when did rock n roll begin?  \end{tabular} 	
    &  \begin{tabular}[l]{p{0.19\textwidth}} when did rock and roll begin \end{tabular} \\ 
    \bottomrule
    \end{tabular}
    \end{table}

%% file: resample_method.tex
\section{Resample-based Evaluation}
\label{sec:resampling_evaluation}

Having shown that the test queries are too similar to the training ones on benchmarks, we argue that it is important to consider interpolation and extrapolation during evaluation and propose a resampling method to evaluate the model performance in the two regimes. 
We resample training queries that are similar or dissimilar to the test queries to evaluate interpolation or extrapolation performance, respectively. The proposed resampling method requires neither new test collections nor annotation. It can be directly applied to any Cranfield-like dataset that consists of a training set and a test collection. We first elaborate on the method and then validate its efficacy by investigating whether the results align with the generalization performance evaluated by new out-of-distribution test sets. 

\subsection{Methodology}

As introduced in Section~\ref{sec:define_interpolation_extrapolation}, we consider the dynamics of queries and define interpolation and extrapolation based on query similarity. Interpolation occurs when the test query is similar to at least one training query. Otherwise, extrapolation occurs. In the following, we will first introduce how we compute query similarity and then utilize the similarity to resample the dataset. 

\subsubsection{Query Similarity} \mbox{}

There are two options for efficiently computing query similarity, i.e., term-based methods~\cite{bm25} and embedding-based retrieval~\cite{zhan2021optimizing}. We utilize the manual annotation results to decide which is better. Recall that Section~\ref{sec:test_train_overlap} annotates similarities between test queries and recalled training queries. A better model should recall more similar queries, resulting in higher annotated similarity scores. According to the annotation results, the embedding-based method recalls more similar queries. Hence it is used to compute query similarity when we resample the dataset. 

\input{latex_figures/resample_method.tex}

\subsubsection{Resampling Training Queries} \mbox{}

The first method is ReSampling Training Queries~(ReSTrain). ReSTrain keeps the original test set unchanged. It selects similar or dissimilar training queries as a new training set for interpolation or extrapolation evaluation, respectively. Concretely, ReSTrain collects top-$I$ similar training queries of each test query as the interpolation training set. It excludes top-$E$ similar training queries of each test query to construct the extrapolation training set. We can tune $I$ and $E$ to explore the influence of training set size. 
We separately train ranking models on the new training sets and evaluate them on the original test set. The gap in the evaluation scores reveals the performance loss when the model transitions from interpolation to extrapolation.

We visualize ReSTrain in Figure~\ref{fig:restrain}. To enable clear visualization, we first use PCA to reduce dimension to two and then resample training queries. 
Blue and orange points indicate extrapolation and interpolation training queries, respectively. As the figure shows, memorization may help improve interpolation performance because the interpolation training queries are close to the test queries. But it hardly helps extrapolation due to the substantial distinction between training and test queries. Thus, models that do not rely on memorization can achieve similar effectiveness in the two evaluation regimes. 

ReSTrain works well for small test sets like TREC 19 \& 20 DL because it can find many distinct training queries to construct a new extrapolation training set. 
Nevertheless, if the test set is large (e.g., MS MARCO dev set), the test queries cover a broad range of topics and thus relate to too many training queries. In this scenario, ReSTrain struggles to evaluate extrapolation performance because only a few truly dissimilar training queries would remain in the extrapolation training set. 
Next, we will introduce another resampling method to address this limitation.

\subsubsection{Resampling Training and Test Queries} \mbox{}

The second evaluation method is ReSampling Training and Test Queries~(ReSTTest), which is specially used for large test sets.  
ReSTTest clusters the large test set to several small ones so that we can evaluate extrapolation performance for a small proportion of the test queries each time. 

Specifically, ReSTTest clusters training queries and test queries into $k$ buckets with k-means. We use training queries in $k-1$ buckets as a new training set. The test queries in those $k-1$ buckets are used for interpolation evaluation, and the test queries in the one remaining bucket are used for extrapolation evaluation. We train and evaluate models $k$ times to acquire full extrapolation results of all test queries. As for interpolation, we report the average evaluation metric score since each test query is judged $k-1$ times. In this paper, $k$ is set to $5$. 

We visualize ReSTTest in Figure~\ref{fig:resttest}. Like Figure~\ref{fig:restrain}, we also use PCA for dimension reduction to obtain clear visualization. 
As shown in the figure, we divide the vector space into five~($k=5$) parts. Each time we select one part. The training queries in the selected part are removed when we train the ranking models.
Then, we use the test queries in this part to evaluate the extrapolation capacity and the remaining test queries to evaluate the interpolation performance.  
We repeat this process five times until all test queries have been used in the extrapolation regime. 


\input{alignment_with_beir}

%% file: latex_figures/resample_method.tex
\begin{figure}[t]
	\subfloat[ReSTrain keeps the original test set unchanged and samples similar/dissimilar training queries for interpolation/extrapolation evaluation.]{\label{fig:restrain}\includegraphics[width=0.46\linewidth]{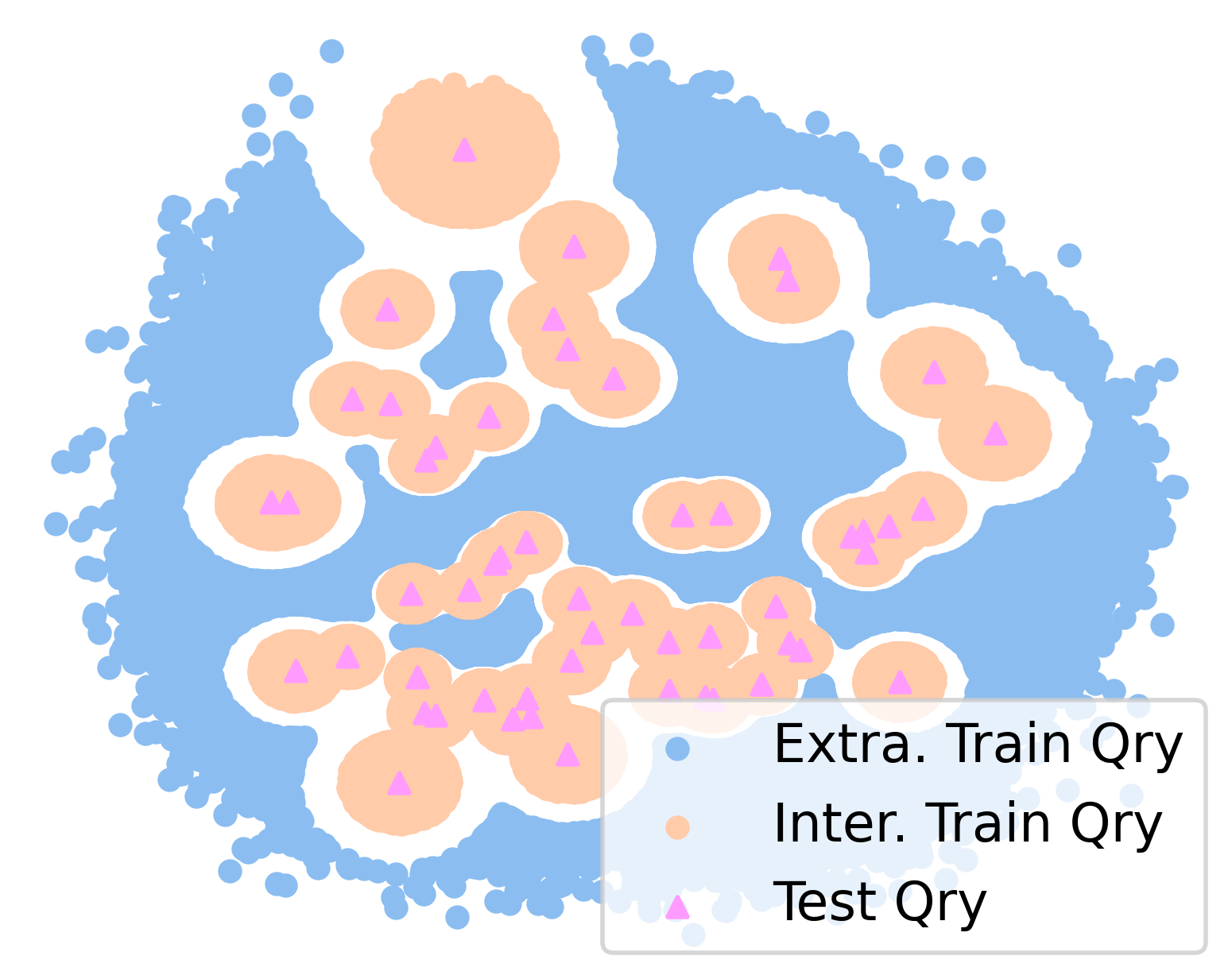}}   
	\hspace{5mm}
	\subfloat[ReSTTest clusters the queries to $k$~(e.g. 5) buckets and selects the test queries in one bucket for extrapolation evaluation. The queries in other buckets are used for training or interpolation evaluation.]{\label{fig:resttest}\includegraphics[width=0.46\linewidth]{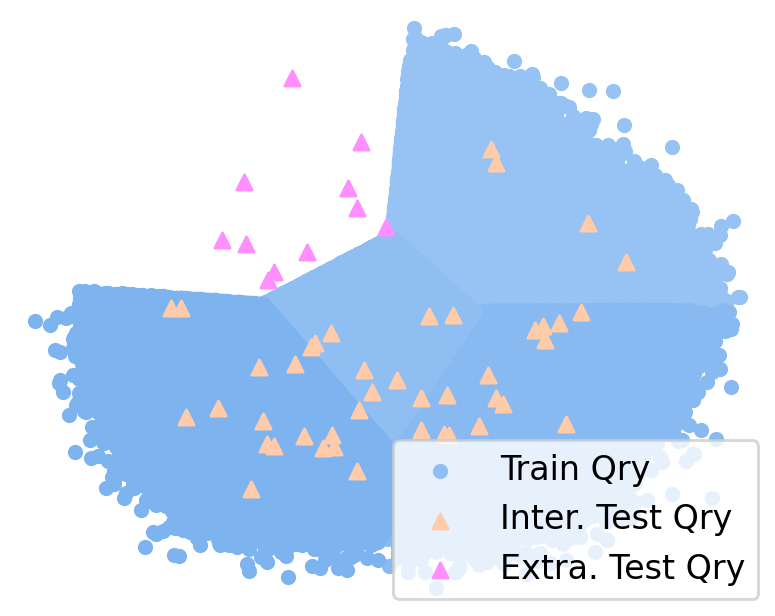}}  
	\caption{
	Schematic diagrams of ReSTrain and ReSTTest. }
\end{figure}

%% file: alignment_with_beir.tex
\subsection{Alignment with Generalization}
\label{sec:align_with_beir}

Evaluation of interpolation and extrapolation should align with generalization performance~\cite{zeni2022exploring}: extrapolation capacity aligns with how well models perform on out-of-distribution test data while interpolation capacity does not. Although we propose an efficient evaluation method that does not require any new testbeds or annotation, the method is limited to a particular distribution that the original dataset follows. It remains to be validated about its alignment with performance on new test sets that follow a different distribution. 
Therefore, we investigate the alignment between our evaluation results and out-of-distribution generalization performance. We first resample the dataset and train multiple variants of neural retrieval models. We then evaluate their interpolation, extrapolation, and out-of-distribution performance. Finally, we compute the performance correlation to investigate the alignment. 

\subsubsection{Implementation Details} \mbox{}

To save the computational costs of training many models, we opt for ReSTrain in this section because it is much more efficient than ReSTTest. The correlation results also generalize to ReSTTest because both methods resample queries based on embedding-based similarity. 
We utilize ReSTrain to resample TREC 19 \& 20 DL passage ranking dataset. We train models and evaluate the interpolation and extrapolation performance.
We resample three sizes of training sets, i.e., 14k, 45k, and 200k training queries. For each size, we resample two training sets, one for interpolation and one for extrapolation.
We use BEIR dataset~\cite{thakur2021beir} to evaluate the zero-shot out-of-distribution performance. It contains various test sets from different domains.
Because different neural architectures are affected differently due to the incomplete annotation of BEIR~\cite{thakur2021beir}, we only use dense retrieval models to investigate the performance correlation to alleviate the problem. We follow the common practice to implement the dense retrieval model~\cite{zhan2021optimizing, xiong2021approximate, karpukhin2020dense}, i.e., feeding the model with queries and passages separately and utilizing the output vector of [cls] token as text embedding. 
We vary the training settings. We use random negative sampling and different pretrained language models for initialization including RoBERTa~\cite{liu2019roberta}, BERT~\cite{devlin2019bert}, ERNIE2.0~\cite{Sun2020ERNIE2A}, Condenser~\cite{gao2021condenser}, coCondenser-wiki~\cite{gao2021unsupervised}, and coCondenser-marco~\cite{gao2021unsupervised}. We also initialize the model with BERT~\cite{devlin2019bert} and utilize advanced training techniques such as hard negative mining~\cite{zhan2021optimizing} and distillation~\cite{lin2020distilling}. Overall, there are $8$ training methods and $3$ training set sizes. We train $24$ models on the training sets for interpolation and another $24$ models on the training sets for extrapolation. We evaluate the out-of-distribution performance of all $48$ models on BEIR.

    
\begin{table}[t]
    \small
    \caption{Performance (NDCG@10) correlation between TREC-Covid and TREC DL. Better correlated results are marked bold except for very low correlation numbers. 
    }
    \label{tab:correlation_coefficient}
    \begin{tabular}{ll|cc|cc}
    \toprule
    \multirow{2}{*}{\textbf{Datasets}} & \multirow{2}{*}{\textbf{Domain}} & \multicolumn{2}{c|}{\textbf{Spearman}} & \multicolumn{2}{c}{\textbf{Kendall}} \\
    & & \textbf{Inter.} & \textbf{Extra.} & \textbf{Inter.} & \textbf{Extra.} \\ \midrule
    {TREC-Covid} 	& Bio-Medical	& 0.537 & \textbf{0.841} & 0.359 & \textbf{0.645} \\
    {NFCorpus} 		& Bio-Medical 	& 0.582 & \textbf{0.783} & 0.418 & \textbf{0.611} \\
    {NQ} 			& Wiki-QA		& 0.791 & \textbf{0.935} & 0.606 & \textbf{0.805} \\
    {HotpotQA} 		& Wiki-QA		& 0.579 & \textbf{0.836} & 0.416 & \textbf{0.635} \\
    {DBPedia} 		& Wiki-Entity	& 0.647 & \textbf{0.846} & 0.492 & \textbf{0.679} \\
    {CQADupS.}		& StackEx. Query& 0.577 & \textbf{0.902} & 0.403 & \textbf{0.730} \\
    {Quora} 		& Quora	Query	& 0.655 & \textbf{0.823} & 0.493 & \textbf{0.647} \\
    {SciDocs} 		& Sci Citation	& 0.595 & \textbf{0.724} & 0.440 & \textbf{0.546} \\
    {SciFact} 		& Sci Citation  & 0.603 & \textbf{0.793} & 0.453 & \textbf{0.611} \\
    {Arguana} 		& Counter-argu.	& \textbf{0.612} & 0.504 & \textbf{0.435} & 0.375 \\
	{FiQA} 			& Finance-QA	& 0.273 & 0.283 & 0.223 & 0.199 \\
	{FEVER} 		& Wiki Fact-check& -0.284 & -0.175 & -0.179 & -0.142 \\
	{Climate-F.} 	& Wiki Fact-check& -0.515 & -0.352 & -0.419 & -0.260 \\
	{Touch{\'e}-2020} 	& Argument		& -0.307 & -0.410 & -0.216 & -0.278 \\
    \midrule
    \bottomrule
    \end{tabular}
    \end{table}

\begin{figure}[t]
	\subfloat[TREC-Covid vs. TREC DL interpolation]{\label{fig:correlation_inter_ndcg}\includegraphics[width=0.45\linewidth]{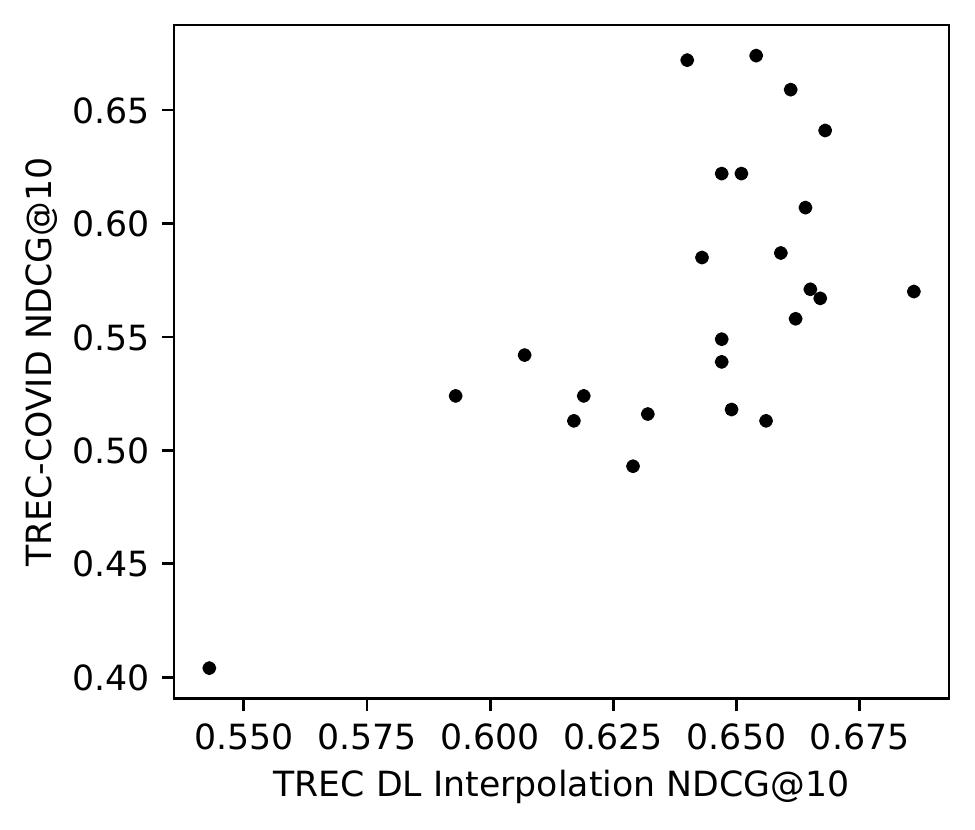}}   
	\hspace{5mm}
	\subfloat[TREC-Covid vs. TREC DL extrapolation]{\label{fig:correlation_extra_ndcg}\includegraphics[width=0.45\linewidth]{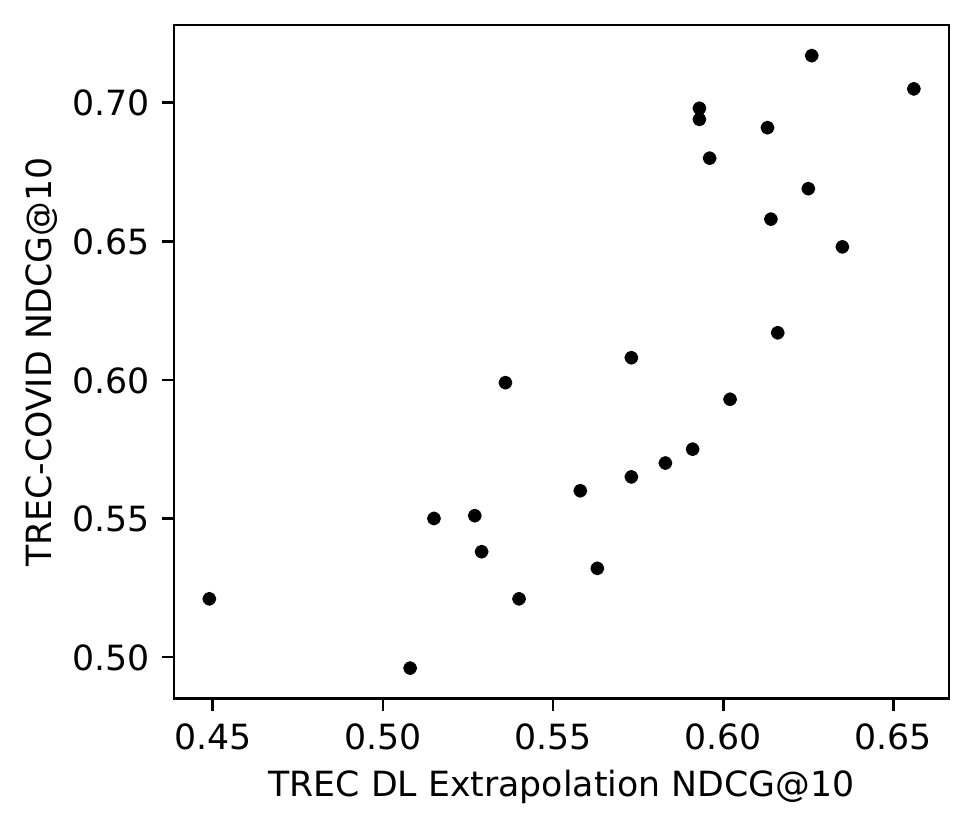}}  
	\caption{Performance~(NDCG@10) on TREC-Covid~(y-axis) and TREC DL~(x-axis). 
	}
	\label{fig:correlation_ndcg}
\end{figure}

\subsubsection{Discussion} \mbox{}

Performance correlation results are shown in Table~\ref{tab:correlation_coefficient}.
We can see that extrapolation performance evaluated on a single benchmark dataset strongly correlates with the performance on most out-of-distribution test sets. On the contrary, interpolation performance is substantially less correlated. 
To present the correlation more clearly, we plot the interpolation and extrapolation correlation in Figures~\ref{fig:correlation_inter_ndcg} and \ref{fig:correlation_extra_ndcg}, respectively. The x-axis shows the performance on the TREC 19 \& 20 DL, and the y-axis shows the out-of-distribution performance on TREC-Covid which is regarded as the most reliable test set in BEIR~\cite{xin2021zero, thakur2021beir}. 
The two figures are consistent with the correlation results in Table~\ref{tab:correlation_coefficient}. The points in Figure~\ref{fig:correlation_extra_ndcg}  clearly follow an upward trend, whereas the points in Figure~\ref{fig:correlation_inter_ndcg} do not. 

We also notice unexpected results in the correlation analysis for some BEIR datasets in Table~\ref{tab:correlation_coefficient}, such as Arguana~\cite{wachsmuth2018retrieval}, FiQA~\cite{maia2018WWW18OC}, FEVER~\cite{thorne2018fever}, Climate-FEVER~\cite{diggelmann2020climate}, and Touch{\'e}-2020~\cite{bondarenko2020OverviewOT}. 
Except for FiQA~\cite{maia2018WWW18OC} whose paper contains little information about the dataset, we find the rest datasets define relevance differently from typical ad-hoc search. 
Concretely, Arguana~\cite{wachsmuth2018retrieval} and Touch{\'e}-2020~\cite{bondarenko2020OverviewOT} are counterargument retrieval and conversational argument retrieval tasks, respectively. 
FEVER~\cite{thorne2018fever} and Climate-FEVER~\cite{diggelmann2020climate} require retrieving evidence that supports given claims.
We conjecture that the inconsistent definitions of relevance harm generalization and induce unexpected correlation results.

Overall, the results validate the efficacy of our proposed method. Even if the proposed method is limited to the distribution that the original dataset follows, resampling the dataset based on query similarity still reliably evaluates the extrapolation performance that aligns with the out-of-distribution retrieval effectiveness. Such a result is meaningful because annotating a new test set is hard and costly.
Besides, the proposed method additionally evaluates the interpolation capacity, which reflects model behavior in an opposite perspective to extrapolation.


%% file: cmp_architecture.tex
\section{Interpolation and Extrapolation Evaluation of Existing Retrieval Models}
\label{sec:evaluate_extrapolation}

Now we investigate the necessity of interpolation and extrapolation evaluation by studying whether evaluation in the two regimes induces different model comparison outcomes. 
We re-evaluate popular neural retrieval models including various model architectures and training methods.  

\subsection{Impact of Model Architectures}
\label{sec:compare_model_architecture}

We compare interaction-based models and representation-based models including dense retrieval and neural sparse retrieval. 
We first introduce the three model architectures, then brief the implementation details, and finally discuss the evaluation results.  

\subsubsection{Model Architectures} \mbox{}

Interaction-based methods model the term-term interactions to predict relevance. For example, BERT reranker~\cite{nogueira2019passage} takes the input of both the query and the passage and computes the attention scores between each term pair to contextualize the representations. ColBERT~\cite{Khattab2020ColBERTEA} is based on BERT and further leverages a late-interaction mechanism to speed up inference. 
We use ColBERT~\cite{Khattab2020ColBERTEA} and BERT reranker~\cite{nogueira2019passage} to represent interaction-based models. 

Representation-based methods involve dense retrieval and neural sparse retrieval. Dense retrieval takes the input of queries and passages separately and outputs their representations. 
Existing dense retrieval models usually share the same architecture, i.e., transformer-based encoder and using the output vector of [cls] as text representation. 
Neural sparse retrieval predicts term weights and builds inverted indexes for efficient retrieval. It can be regarded as encoding text to a sparse and vocabulary-sized vector. We select the popular SPLADE~\cite{formal2021splade} model as a representative.  


\input{latex_tables/restrain_architecture_compare.tex}

\subsubsection{Implementation} \mbox{}

We use the authors' open-sourced code for training ColBERT~\cite{Khattab2020ColBERTEA} and SPLADE~\cite{formal2021splade}. We implement BERT reranker by using pairwise cross-entropy loss. 
As for dense retrieval, we adopt a base training setting, i.e., $1,024$ random negatives per mini-batch and cross-entropy loss. 
We denote the trained dense retrieval model as DR~(base) and will explore other advanced training methods in the next section. 
All neural models are initialized with BERT-base-uncased~\cite{devlin2019bert} and are trained for $50$ epochs. 
For evaluation settings, BERT reranker reranks the top-1000 passages retrieved by BM25, and SPLADE, ColBERT, and dense retrieval end-to-end retrieve top passages from the entire corpus.
We evaluate the checkpoint every 2 epochs and select the best one according to the performance on the development set.
Besides the above neural models, we involve BM25~\cite{yang2018anserini} as a traditional baseline and tune its parameters on the training data with grid search. 

\input{latex_tables/resttest_architecture_compare.tex}

\input{latex_tables/compare_hard_distill_pretrain.tex}

We utilize ReSTrain to evaluate interpolation and extrapolation performance on TREC DL test sets. 
We set the number of training queries to 14k, 45k, and 200k to explore the influence of training data. The results are shown in Table~\ref{tab:restrain_architecture_compare}.
We use ReSTTest to investigate the interpolation and extrapolation performance on the large MS MARCO dev set. All training and test queries are clustered into five buckets. The results are shown in Table~\ref{tab:resttest_architecture_compare}.

\subsubsection{Discussion} \mbox{}

According to Tables~\ref{tab:restrain_architecture_compare} and \ref{tab:resttest_architecture_compare}, we can see that representation-based models, including SPLADE and dense retrieval, suffer from a substantial loss in ranking performance when it transitions from interpolation to extrapolation. Take dense retrieval for example. On TREC DL test sets in Table~\ref{tab:restrain_architecture_compare}, the NDCG@10 values decrease more than $10\%$, and R@100 numbers decrease about $8\%$. The trends are similar on the MSMARCO Dev set in Table~\ref{tab:resttest_architecture_compare}. On the contrary, interaction-based models are much more robust to extrapolation. BERT reranker and ColBERT extrapolate almost as well as they interpolate in terms of both metrics on TREC DL (Table~\ref{tab:restrain_architecture_compare}) and R@100 on MSMARCO Dev set (Table~\ref{tab:resttest_architecture_compare}).
Therefore, representation-based models seem to heavily rely on interpolation and are vulnerable to extrapolation. Interaction-based models benefit from modeling term-term interactions and extrapolate well to novel test queries.

Results also suggest that dense retrieval and SPLADE are able to interpolate even similarly to interaction-based models but extrapolate substantially worse. On TREC DL test sets, dense retrieval and SPLADE perform comparably with ColBERT when trained on the interpolation training set. But they substantially underperform when extrapolating. On the MSMARCO Dev set, dense retrieval and SPLADE even outperform the BM25-BERT reranking system in terms of Recall by a large margin when they interpolate. But during extrapolation evaluation, BM25-BERT performs better. 

Therefore, interpolation evaluation and extrapolation evaluation may draw different conclusions about which model is better. In a constrained scenario where all possible queries are limited, representation-based models excel because of good interpolation ability and high efficiency. Instead, if the models are required to be applied in search engines where queries are dynamically changing, interaction-based models excel due to strong extrapolation capacity. 

%% file: latex_tables/restrain_architecture_compare.tex
\begin{table}[t]
    \small
    \caption{Interpolation and extrapolation performance on TREC 19 \& 20 DL evaluated with ReSTrain. $\Delta$=Extra./Inter.-1. \#Q denotes the number of training queries. Reranker and ColBERT are interaction-based. SPLADE and dense retrieval~(DR) are representation-based.}
    \label{tab:restrain_architecture_compare}
    \begin{tabular}{cl|ccr|ccr}
    \toprule
    \multirow{3}{*}{\textbf{\#Q}} & \multirow{3}{*}{\textbf{Models}} & \multicolumn{6}{c}{\textbf{TREC 19 \& 20 DL~(ReSTrain)}}  \\
    & & \multicolumn{3}{c|}{\textbf{NDCG@10}} & \multicolumn{3}{c}{\textbf{R@100}} \\
    & & {\textbf{Inter.}} & {\textbf{Extra.}} & \multicolumn{1}{c|}{$\Delta$} & {\textbf{Inter.}} & {\textbf{Extra.}} & \mc{$\Delta$} \\
    \midrule
    14k & BM25 & 0.485 & 0.490 & +1\% & 0.534 & 0.541 & +1\% \\
    \midrule
    \multirow{3}{*}{14k}
      & Reranker	& 0.679 & 0.645 & -5\% & 0.645 & 0.639 & -1\% \\
      & ColBERT 	& 0.597 & 0.585 & -2\% & 0.567 & 0.605 & +7\% \\
      & SPLADE	 	& 0.596 & 0.488 & -18\% & 0.588 & 0.547 & -7\% \\
      & DR~(base)	& 0.617 & 0.529 & -14\% & 0.576 & 0.534 & -7\% \\
    \midrule
    \multirow{3}{*}{45k}
      & Reranker	& 0.692 & 0.680 & -2\% & 0.659 & 0.650 & -1\% \\
      & ColBERT 	& 0.649 & 0.633 & -2\% & 0.612 & 0.633 & +3\% \\
      & SPLADE	 	& 0.615 & 0.512 & -17\% & 0.615 & 0.567 & -8\% \\
      & DR~(base) 	& 0.647 & 0.558 & -14\% & 0.607 & 0.558 & -8\% \\
    \midrule
    \multirow{3}{*}{200k}
      & Reranker	& 0.709 & 0.705 & -1\% & 0.660 & 0.664 & +1\% \\
      & ColBERT 	& 0.676 & 0.661 & -2\% & 0.633 & 0.649 & +3\% \\
      & SPLADE 		& 0.637 & 0.607 & -5\% & 0.642 & 0.611 & -5\% \\
      & DR~(base) 	& 0.664 & 0.591 & -11\% & 0.621 & 0.564 & -9\% \\
    \bottomrule
    \end{tabular}
    \end{table}

%% file: latex_tables/resttest_architecture_compare.tex
\begin{table}[t]
    \small
    \caption{Interpolation and extrapolation performance on MS MARCO evaluated with ReSTTest. $\Delta$=Extra./Inter.-1. Reranker and ColBERT are interaction-based. SPLADE and dense retrieval~(DR) are representation-based.}
    \label{tab:resttest_architecture_compare}
    \begin{tabular}{l|ccr|ccr}
    \toprule
    \multirow{3}{*}{\textbf{Models}} & \multicolumn{6}{c}{\textbf{MSMARCO Dev~(ReSTTest)}}  \\ 
    & \multicolumn{3}{c|}{\textbf{MRR@10}} & \multicolumn{3}{c}{\textbf{R@100}} \\
    & {\textbf{Inter.}} & {\textbf{Extra.}} & \multicolumn{1}{c|}{$\Delta$} & {\textbf{Inter.}} & {\textbf{Extra.}} & \mc{$\Delta$} \\
    \midrule
    BM25 		& 0.189 & 0.188 & -1\% & 0.671 & 0.669 & 0\% \\
    \midrule
    Reranker	& 0.369 & 0.346 & -6\% & 0.824 & 0.815 & -1\% \\
    ColBERT 	& 0.358 & 0.324 & -9\% & 0.877 & 0.848 & -3\% \\
    SPLADE 		& 0.324 & 0.282 & -13\% & 0.867 & 0.812 & -6\% \\
    DR~(base)	& 0.313 & 0.277 & -12\% & 0.848 & 0.789 & -7\% \\
    \bottomrule
    \end{tabular}
    \end{table}

%% file: latex_tables/compare_hard_distill_pretrain.tex
\begin{table*}[t]
    \small
    \caption{Interpolation and extrapolation ranking performance of dense retrieval models~(DR) trained with different methods. $\Delta$=Extra./Inter.-1. The training set size for ReSTrain is 45k.}
    \label{tab:compare_hard_distill_pretrain}
    \begin{tabular}{lll|ccr|ccr|ccr|ccr}
    \toprule
    \multirow{3}{*}{\textbf{Type}} &
    \multirow{3}{*}{\textbf{PLM}} & \multirow{3}{*}{\textbf{Finetune}} & \multicolumn{6}{c|}{\textbf{TREC 19 \& 20 DL~(ReSTrain)}} & \multicolumn{6}{c}{\textbf{MS MARCO Dev~(ReSTTest)}}  \\ 
    & & & \multicolumn{3}{c|}{\textbf{NDCG@10}} & \multicolumn{3}{c|}{\textbf{R@100}} & \multicolumn{3}{c|}{\textbf{MRR@10}} & \multicolumn{3}{c}{\textbf{R@100}} \\
    & & & {\textbf{Inter.}} & {\textbf{Extra.}} & \multicolumn{1}{c|}{$\Delta$} & {\textbf{Inter.}} & {\textbf{Extra.}} & \multicolumn{1}{c|}{$\Delta$} & {\textbf{Inter.}} & {\textbf{Extra.}} & \multicolumn{1}{c|}{$\Delta$} & {\textbf{Inter.}} & {\textbf{Extra.}} & \multicolumn{1}{c}{$\Delta$} \\
    \midrule
    DR~(base) & BERT & RandNeg 		& 0.647 & 0.558 & -14\% & 0.607 & 0.558 & -8\% & 0.313 & 0.277 & -12\% & 0.848 & 0.789 & -7\% \\
    \midrule
    \multirow{3}{*}{DR~(advance)}
    & BERT 		& HardNeg	& 0.661 & 0.581 & -12\% & 0.620 & 0.558 & -10\% & 0.334 & 0.289 & -13\% & 0.866 & 0.801 & -8\% \\
    & BERT 		& Distill 	& 0.656 & 0.573 & -13\% & 0.586 & 0.554 & -5\% & 0.329 & 0.291 & -12\% & 0.852 & 0.797 & -6\% \\
    & coCon-marco 	& RandNeg	& 0.668 & 0.626 & -6\% & 0.641 & 0.644 & 0\% & 0.323 & 0.299 & -7\% & 0.875 & 0.850 & -3\% \\
    \bottomrule
    \end{tabular}
    \end{table*}

%% file: cmp_dr_train.tex
\subsection{Impact of Training Strategies}
\label{sec:compare_dr_train}

Having shown that different model architectures exhibit different interpolation and extrapolation capabilities, we now re-evaluate various training techniques to explore whether their effectiveness also differs in the two regimes.
We consider hard negative mining~\cite{xiong2021approximate, zhan2021optimizing}, distillation~\cite{lin2020distilling, hofstatter2021efficiently}, and pretraining~\cite{gao2021condenser, gao2021unsupervised}. They are originally proposed to improve effectiveness of dense retrieval models and are also demonstrated to be effective for interaction-based models~\cite{santhanam2021colbertv2} and neural sparse retrieval~\cite{formal2021spladev2}. In this paper, we evaluate their efficacy based on dense retrieval models. 

\subsubsection{Advanced Training Methods} \mbox{}

There are mainly three methods to improve the efficacy of dense retrieval. 
First, \citet{xiong2021approximate} and \citet{zhan2021optimizing} employ hard negatives during training. \citet{xiong2021approximate} asynchronously sample hard negatives, and \citet{zhan2021optimizing} further propose synchronous sampling via freezing the document index. 
Second, \citet{hofstatter2021efficiently} and \citet{lin2020distilling} propose to distill BERT rerankers and ColBERT, respectively. The two interaction-based models exhibit strong ranking performance and thus are suitable teachers for dense retrieval. Distilling ColBERT is more computationally friendly because ColBERT is more efficient than BERT reranker.
Finally, pretraining methods tailored for dense retrieval are also effective~\cite{gao2021condenser, gao2021unsupervised}. Condenser~\cite{gao2021condenser} utilizes an auto-encoding task, and coCondenser~\cite{gao2021unsupervised} leverages contrastive learning.

\subsubsection{Implementation} \mbox{}

We follow STAR and ADORE~\cite{zhan2021optimizing} to implement hard negative mining. And we follow TCT-ColBERT~\cite{lin2020distilling} to implement distillation. The metric scores when we use the full training set match the numbers reported in previous studies, demonstrating the correctness of our implementation. As for pretraining techniques, we explore a variety of pretrained language models~(PLM), including BERT~\cite{devlin2019bert}, RoBERTa~\cite{liu2019roberta}, ERNIE 2.0~\cite{Sun2020ERNIE2A}, Condenser~\cite{gao2021condenser}, and coCondenser~\cite{gao2021unsupervised}. All models except coCondenser are trained on Wikipedia corpus. coCondenser has two variants. One denoted as coCon-wiki is pretrained on Wikipedia. The other denoted as coCon-marco is pretrained on Wikipedia and MS MARCO.

\subsubsection{Comparing Pretraining and Finetuning} \mbox{}

This section investigates how pretraining and finetuning help improve interpolation and extrapolation. The results are shown in Table~\ref{tab:compare_hard_distill_pretrain}. We can see that initializing models with coCon-marco~\cite{gao2021condenser} is the most effective way to improve extrapolation capacity. It substantially reduces the interpolation-extrapolation performance gap, whereas distillation~\cite{lin2020distilling} and hard negative mining~\cite{zhan2021optimizing} cannot. It also remarkably improves the absolute extrapolation performance. For example, measured by R@100, its extrapolation performance nearly matches or even outperforms the other models' interpolation performance. 
On the contrary, hard negative mining and distillation are effective for interpolation. For example, they interpolate better than coCon-marco in MRR@10 on the MS MARCO dev set.  

Therefore, the results imply that pretraining is more effective than finetuning in improving the extrapolation ability. It again verifies the importance to evaluate interpolation and extrapolation performance separately. 
Considering the impressive performance of coCon-marco, we decide to perform a deep dive into how pretraining helps extrapolation.

\subsubsection{Investigation of Pretraining} \mbox{}
\label{sec:compare_dr_train_invest_pretrain}

We comprehensively study the efficacy of pretraining to gain insights into what pretraining techniques are effective for improving extrapolation performance.
We initialize encoders with various pretrained language models~(PLM) and then finetune them using the base training setting, i.e., $1024$ random negatives per training step and cross-entropy loss. We also directly finetune a randomly initialized model as a baseline. The results are shown in Table~\ref{tab:compare_pretrain}. 

\input{latex_tables/compare_pretrain.tex}

First, the results highlight the importance of pretrained language models. Random initialization leads to $30\%$ and $16\%$ performance gap on TREC DL and MS MARCO dev set, respectively, which is about twice the gap when we use RoBERTa~\cite{liu2019roberta} for initialization. Table~\ref{tab:visualize_pretrain} visualizes the passage representations of models with and without pretraining.
 Different colors indicate relevant passages for different TREC 20 DL test queries. We can see that a finetuned random-initialized model~(Rand) clusters the passage representations poorly and appropriately in the extrapolation and interpolation regimes, respectively. Instead, the pretrained language model~(coCon-marco) is already equipped with the representation ability and gets better clustering results than the finetuned random-initialized model. It also substantially reduces the gap between interpolation and extrapolation as the clustering results barely change when moving from interpolation to extrapolation.

\input{latex_tables/visualize_pretrain.tex}

Second, it is important to adopt a text-encoding pretraining task. RoBERTa does not use such a task and only pretrains the model on word level. It achieves the worst extrapolation capacity among all pretrained language models. Others adopt different text-encoding tasks, such as next sentence prediction~(BERT), sentence ordering prediction~(ERNIE2.0), auto-encoding~(Condenser), and contrastive learning~(coCon-wiki \& coCon-marco). 

Third, pretraining on the target corpus is beneficial. Both coCon-marco and coCon-wiki adopt contrastive learning for pretraining, while coCon-marco is additionally pretrained on the target corpus and extrapolates substantially better. 
However, pretraining on the target corpus may be less effective if the definitions of interpolation and extrapolation further consider the dynamics of documents, i.e., some documents are not available during pretraining or finetuning. We will explore it in the future.

Overall, the separation of interpolation and extrapolation helps deeply understand the effectiveness of different methods. Results show that finetuning is beneficial for interpolation and suitable pretraining techniques have a pronounced effect on improving extrapolation capacity. 

%
%

%% file: latex_tables/compare_pretrain.tex
\begin{table}[t]
    \small
    \caption{Ranking performance of dense retrieval models when initialized by different PLMs. $\Delta$=Extra./Inter.-1. The training set size for ReSTrain is 45k. TE and TC stand for text-encoding and target-corpus pretraining, respectively.}
    \label{tab:compare_pretrain}
    \begin{tabular}{p{16mm}C{1.5mm}C{2.5mm}|C{5mm}C{5mm}R{6mm}|C{5mm}C{5mm}R{6mm}}
    \toprule
    \multirow{3}{*}{\textbf{PLM}} & \multirow{3}{*}{\textbf{TE}} & \multirow{3}{*}{\textbf{TC}} & \multicolumn{3}{c|}{\textbf{TREC 19\&20 DL}} & \multicolumn{3}{c}{\textbf{MSMARCO Dev}}  \\ 
    & & & \multicolumn{3}{c|}{\textbf{R@100~(ReSTrain)}} & \multicolumn{3}{c}{\textbf{R@100~(ReSTTest)}} \\
    & & & {\textbf{Inter.}} & {\textbf{Extra.}} & \multicolumn{1}{c|}{$\Delta$} & {\textbf{Inter.}} & {\textbf{Extra.}} & \multicolumn{1}{c}{$\Delta$} \\
    \midrule
     Rand			& &	& 0.437 & 0.305 & -30\% & 0.723 & 0.606 & -16\% \\
     RoBERTa		& &		& 0.582 & 0.499 & -14\% & 0.845 & 0.778 & -8\% \\
     BERT			& \checkmark &		& 0.607 & 0.558 & -8\% & 0.848 & 0.789 & -7\% \\
     ERNIE2.0		& \checkmark &		& 0.610 & 0.568 & -7\% & 0.856 & 0.812 & -5\% \\
     Condenser		& \checkmark &		& 0.615 & 0.580 & -6\% & 0.859 & 0.811 & -6\% \\
     coCon-wiki		& \checkmark &		& 0.615 & 0.582 & -5\% & 0.862 & 0.819 & -5\% \\
     coCon-marco	& \checkmark &	\checkmark & 0.641 & 0.644 & 0\% & 0.875 & 0.850 & -3\% \\
    \bottomrule
    \end{tabular}
    \end{table}

%% file: latex_tables/visualize_pretrain.tex
\begin{table}[t]
    \begin{tabular}{ l | c c c }
    \toprule
    \textbf{Init} & \textbf{Zeroshot} & \textbf{Extrapolation} & \textbf{Interpolation} \\ \midrule
    \textbf{Rand} & \raisebox{-.5\height}{\includegraphics[width=0.24\linewidth]{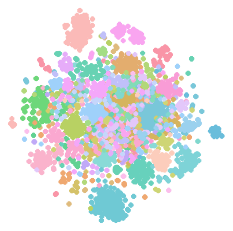}} & \raisebox{-.5\height}{\includegraphics[width=0.24\linewidth]{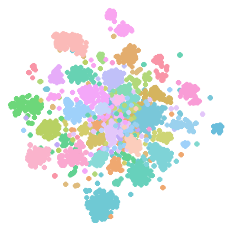}} & \raisebox{-.5\height}{\includegraphics[width=0.24\linewidth]{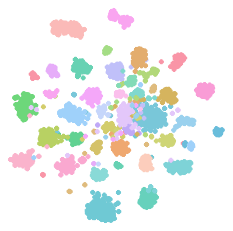}} \\ 
    \textbf{coCon} & \raisebox{-.5\height}{\includegraphics[width=0.24\linewidth]{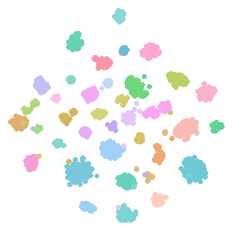}} & \raisebox{-.5\height}{\includegraphics[width=0.24\linewidth]{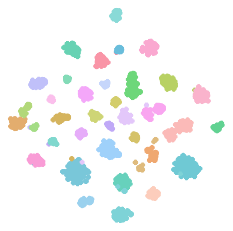}} & \raisebox{-.5\height}{\includegraphics[width=0.24\linewidth]{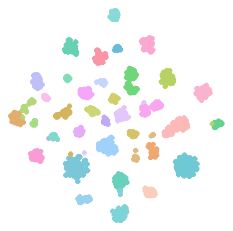}} \\
    \bottomrule
    \end{tabular}
    \caption{Passage representation distribution in different evaluation paradigms of a (finetuned) randomly-initialized model or coCondenser-marco model. Different colors indicate relevant passages for different queries.  
    \label{tab:visualize_pretrain}}
    \end{table}


%% file: related_work.tex
\section{Related Work}
\label{sec:related_work}

We recap related work in neural retrieval and studies that investigate its generalization ability.

Recent efforts in improving the effectiveness of ranking models can be roughly classified into two directions, i.e., model architectures and training techniques. ColBERT~\cite{Khattab2020ColBERTEA, santhanam2021colbertv2}, dense retrieval~\cite{zhan2021optimizing, xiong2021approximate, gao2021unsupervised, lin2020distilling} and neural sparse retrieval~\cite{lin2021few, formal2021spladev2, formal2021splade} are popular model architectures. ColBERT utilizes late-interaction to model the term-term interactions and thus is interaction-based. Neural sparse retrieval and dense retrieval encode queries and documents separately to vectors and utilize vector similarity for search. Unlike ColBERT, they are representation-based. Neural sparse retrieval predicts term weights and outputs a sparse vector of vocabulary size, while dense retrieval embeds text to dense vectors of much lower dimensions. 
Besides innovating model architectures, some researchers propose effective training techniques. 
\citet{xiong2021approximate} and \citet{zhan2021optimizing} highlight the importance of using hard negatives and investigate how to mine them during training.  
\citet{hofstatter2021efficiently} and \citet{lin2020distilling} explore using knowledge distillation methods to train dense retrieval models with expressive yet slow interaction-based models like BERT rerankers~\cite{nogueira2019passage} and ColBERT~\cite{Khattab2020ColBERTEA}. 
\citet{gao2021condenser} and \citet{lu2021less} study pretraining techniques tailored for dense retrieval. 
The three training techniques complement each other and can be combined in practice. 


Recently, several researchers challenge the generalization ability of neural retrieval models by constructing a new test set that follows a different distribution from the training set. 
\citet{sciavolino2021simple} propose a new entity-rich dataset and find dense retrieval performs poorly. \citet{thakur2021beir} introduce BEIR dataset which contains multiple test sets from different domains. 
Although proposing a new out-of-distribution test set enables evaluating generalization performance, annotating such a test set is costly. Usually, the out-of-distribution test set should be in a low-resource domain such as medicine where annotation requires expertise. Furthermore, evaluating neural retrieval models may suffer from `false negative' issue and thus a large annotation pool is required. For example, the BEIR authors note that the annotation is incomplete and that the conclusion may not hold if more items are annotated. 
Different from previous studies that evaluate how models generalize on an out-of-distribution test set, we try to take a closer look at the generalization performance without a new test set by analyzing model behavior in the interpolation and extrapolation regimes. This is done by resampling the dataset where models are trained.

%% file: conclusions.tex
\section{Conclusions}
\label{sec:conclusion}

In this paper, we propose to separately evaluate interpolation and extrapolation capabilities of neural retrieval models. Considering the dynamics of queries in Web search, we define them based on whether the training and test queries are similar or not. Based on the definition, we investigate the bias in popular benchmarks, design associated evaluation methods, and revisit existing neural ranking models. We observe that the popular benchmarks are biased towards interpolation and thus may not reflect how models extrapolate. With the proposed evaluation methods, we find different model architectures and training techniques perform differently in the two regimes. Therefore, it is necessary to consider the two evaluation regimes in future IR studies. 

Our work provides a new perspective to evaluate neural retrieval models in the large data regime. It complements previous IR evaluation studies by further considering the training-test relationship. There is plenty to do in this direction. First, since ranking models should be effective for not only unseen queries but also new documents, dynamics of documents can also be incorporated into the definitions of interpolation and extrapolation. Second, this paper defines interpolation and extrapolation without considering the specific ranking models. It would be interesting to explore whether there should be custom definitions for different models to better align with the generalization performance. Third, optimizing ranking models and evaluating extrapolation performance can be regarded as a min-max game. How to design a ranking model that is robust to extrapolation remains to be investigated.